# Caustics and Rogue Waves in an Optical Sea

Amaury Mathis[1], Luc Froehly[1], Shanti Toenger[1], Frédéric Dias[2],
Goëry Genty[3], John M. Dudley[1]

[1]*Institut FEMTO-ST, UMR 6174 CNRS-Université de Franche-Comté, Besançon, France*
[2]*School of Mathematical Sciences, University College Dublin, Belfield, Dublin 4, Ireland*
[3]*Department of Physics, Tampere University of Technology, Tampere, Finland*

**ABSTRACT**

There are many examples in physics of systems showing *rogue wave* behaviour, the generation of high amplitude events at low probability. Although initially studied in oceanography, rogue waves have now been seen in many other domains, with particular recent interest in optics. Although most studies in optics have focussed on how nonlinearity can drive rogue wave emergence, purely linear effects have also been shown to induce extreme wave amplitudes. In this paper, we report a detailed experimental study of linear rogue waves in an optical system, using a spatial light modulator to impose random phase structure on a coherent optical field. After free space propagation, different random intensity patterns are generated, including partially-developed speckle, a broadband caustic network, and an intermediate pattern with characteristics of both speckle and caustic structures. Intensity peaks satisfying statistical criteria for rogue waves are seen especially in the case of the caustic network, and are associated with broader spatial spectra. In addition, the electric field statistics of the intermediate pattern shows properties of an "optical sea" with near-Gaussian statistics in elevation amplitude, and trough-to-crest statistics that are near-Rayleigh distributed but with an extended tail where a number of rogue wave events are observed.

## I. INTRODUCTION

Rogue waves are statistically rare events with extreme amplitude or intensity which emerge seemingly spontaneously in a particular physical system. Although initially studied in the context of describing the large and destructive waves appearing on the ocean [1-3], the field of rogue wave science has now expanded to include extreme and rare fluctuations in many other systems. The generalisation of the rogue wave concept began with studies of noise-induced soliton wavelength jitter in fibre supercontinuum (SC) generation [4], but since then rogue wave behaviour has been seen in many other physical systems [5]. In the particular field of optics, the optical rogue wave terminology is no longer restricted to solitons in SC generation or even propagation effects in optical fibre. Optical rogue waves have been seen as localised breathers in modulation instability (MI), in optical amplifiers, in instabilities in lasers, and high power pulse filamentation [6], as well as in speckle and other spatial patterns in cavities, multimode fibre, and photorefractive systems [7-11]. In fact, what is generally accepted as the meaning of the terminology of 'rogue wave' is now very broad: a high amplitude event in a system appearing in the long tails of an associated probability distribution which satisfies particular statistical criterion [5,6,12].

Although much attention has been paid to the role of nonlinearity in generating rogue waves [13-17], other results in both oceanography [1-3] and using electromagnetic waves at optical and microwave frequencies [8,18] have reported rogue wave behaviour in purely linear systems. A specific linear mechanism for oceanic rogue wave formation is the concentration or focussing of wave action in a caustic region, and extensive studies have shown how this can lead to a wide variety of rogue wave behaviour [1,2, 19-28]. Of course, although caustics are observed in any wave system, it is perhaps in optics that they are most well-known, associated with a distinct region of high intensity formed from an envelope of light rays reflected or refracted by a curved surface [29,30]. Somewhat surprisingly, however, links between the formation of caustics in optics and the formation of rogue wave events in optical systems have not been the subject of detailed study. The objective of this paper is to address this shortcoming.

In particular, we report here a detailed study of rogue wave statistics in an optical system due to the caustic focussing of a random coherent spatial field. Our experimental setup allows us to conveniently measure intensity statistics based on peak detection over the two dimensional generated spatial pattern, and we show explicitly the presence of intensity peaks satisfying statistical rogue wave criteria. We are also able to correlate the appearance of rogue wave events in the intensity probability distribution with patterns having broader spectra. Our system is conceptually extremely simple, involving only random initial phase and subsequent free space propagation, and the fact that we are so clearly able to see extreme events is particularly striking evidence for the importance of linear effects in generating rogue waves.

## II. SETUP

Our experimental setup is shown schematically in Figure 1. A coherent beam from a continuous wave laser (Helium Neon laser) with λ = 632.8 nm is expanded to fill the aperture of a Hamamatsu LCOS spatial light modulator (SLM), X-10468 series, which has a 600 × 800 pixels array over dimensions of 12 × 16 mm. The SLM then encodes a random two dimensional spatial phase pattern on the beam over a square 600 × 600 pixels array using 256 levels. The phase variations are smoothed over typically 10 pixels so that one can consider the SLM as physically equivalent to a random continuous refracting surface. The subsequent propagation of the beam results in the development of random intensity maxima and minima in the spatial beam profile, following the well-known physics of the development of optical speckle [31-33]. An imaging system is used to reduce the size of the beam after the SLM to the measurement region indicated in Fig. 1 and to fill the aperture of the detector. The imaging system in Fig.1 consisted of four lenses with $f_1$ = 500 mm, $f_2$ = 250 mm, $f_3$ = 100 mm, $f_4$ = 9 mm. The pattern detection system used an IDS uEye UI-3240CP CCD camera with a 1280 × 1024 pixels array over dimensions 6.78 × 5.43 mm and with a 10 bit ADC. Magnification of ×15 was used to fill the CCD camera aperture. Translation of the CCD camera was possible over a 500 µm range from the input plane of the measurement zone using a Newport ILS-LM Series precision translation stage that was displaced in steps of 5 µm. In our experiments, the measurements are not made in order to follow the field evolution to a fully-developed (granular) speckle, but we focus rather on the regime where smoothly-varying random intensity fluctuations are observed.

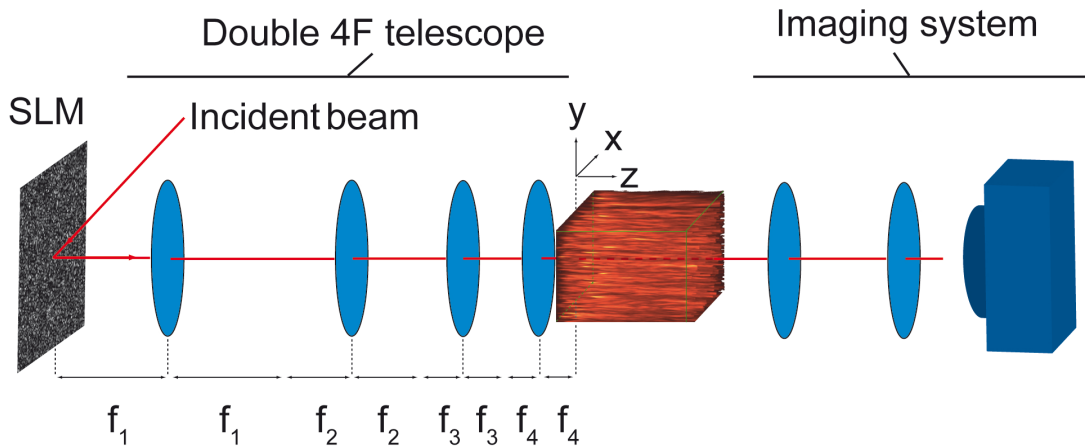

**FIGURE 1** Setup for studying statistics of partially-developed speckle and caustics. An SLM encodes random spatial phase on a coherent beam from a He-Ne laser. Free space propagation transforms this random phase to random intensity fluctuations. An imaging system is used to reduce the size of the beam so it can be recorded on a CCD camera which can be translated longitudinally over an extended measurement volume. All measurement distances given in the text are relative to the origin z = 0 of the axes shown. Here $f_1$ = 500 mm, $f_2$ = 250 mm, $f_3$ = 100 mm, $f_4$ = 9 mm.

## III. NUMERICAL MODELLING

Results showing numerical simulations modelling propagation through the experimental setup are shown in Fig. 2. The simulations discretized the incident field profile to match the experimental SLM pixellation, upon which a smoothed random phase function was applied. A Gaussian fit to the experimental laser beam incident on the SLM was used as the input field. The Angular Spectrum of Plane Waves method [34] was then used to numerically propagate this beam through the optical system described above. No paraxial approximations were made in the modelling.

A coherent field with random phase will initially evolve into a regime of caustic focussing before developing progressively into a granular speckle pattern with propagation distance [33]. The precise nature of the evolution and the structure of the observed pattern depend in detail on the magnitude of the random phase variations, as this determines the random focussing of different portions of the beam at any particular distance. In our simulations and experiments, we choose applied phase distributions in order to highlight: (i) the differences (in form and statistics) between partially-developed speckle and caustic structures; and also (ii) the correlation of high intensity caustics and rogue wave peaks with a broader intensity spectrum. Note that for the differing initial phase distributions, the distances where particular characteristics are observed were determined from our numerical modelling which was used to guide our experiments.

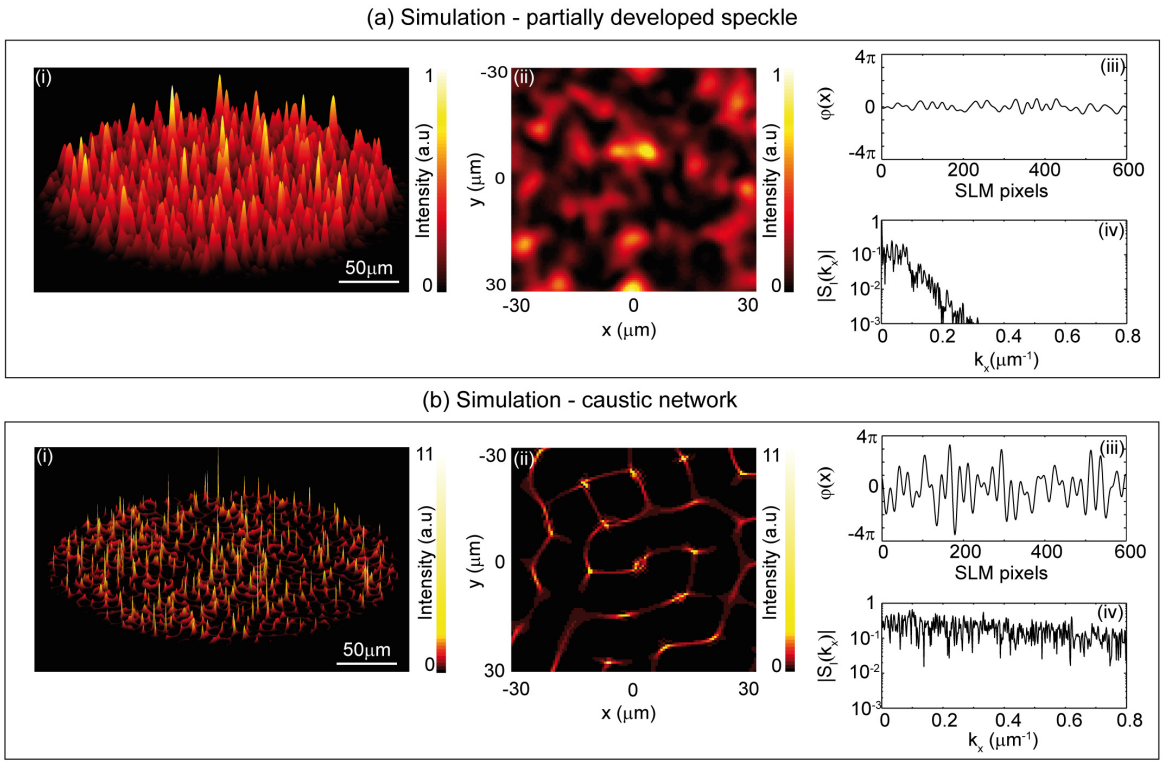

**FIGURE 2** Numerical simulations showing (a) partially-developed speckle and (b) a caustic network. For each case, (i) shows the computed intensity distribution; (ii) shows a zoom over a more limited region looking down on the pattern; (iii) shows a slice of the applied phase distribution to the SLM at y = 0; (iv) shows the calculated spatial spectrum. Intensities in (b) are normalised relative to the maximum intensity for the partially-developed speckle in Fig. 2a. Note the different intensity scales used between (a) and (b).

Figure 2 presents simulation results for two different regimes of evolution. We first show in Fig. 2(a) results generated from an applied phase pattern associated with a narrow spectrum, which leads to a very clear partially-developed speckle at 400 μm. For this applied phase at an earlier stage of evolution at z = 100 μm from the SLM, some caustic network structures were observed, but the strength of focussing was not sufficient to lead to high amplitude peaks and rogue wave statistics. To see rogue wave formation more clearly, it is necessary to use an applied phase pattern associated with a stronger initial random phase modulation and a broader spectrum. These results are shown in Fig. 2(b), where stronger focussing is observed and the caustic network and peaks are seen at a much closer distance to the SLM at z = 10 μm.

For the partially-developed speckle, Fig. 2a(i) plots the computed intensity distribution, with Fig. 2a(ii) showing a zoom over a more limited region looking down on the pattern. Fig. 2a(iii) shows the (unwrapped) applied phase distribution to the SLM (in fact we show a slice $\varphi(x, y = 0)$), and Fig. 2a(iv) shows a slice of the calculated spatial spectrum $S(k_x, k_y = 0)$ which is defined as:

$$S(k_x, k_y) = \left| \int_{-\infty}^{\infty} I(x,y) e^{-i(k_x x + k_y y)} \, dx \, dy \right|$$

where $k_x$ and $k_y$ are the spatial frequencies. Note that we plot results only for positive frequencies as the results are symmetric about $k_x = k_y = 0$. The spatial spectral width (FWHM) from the spectrum is $\Delta k_x \sim 0.1 \, \mu m^{-1}$, consistent with the typical transverse size of the large intensity peaks in the pattern of $\Delta x \sim 8.5 \, \mu m$. These results clearly show the smooth intensity variations typical of a partially-developed speckle.

The results in Fig. 2b(i) and (ii) reveal the characteristic signature of caustics in the form of distinct networks of lines along which light is strongly focussed, as well as particular "hot spots" of intensity peaks where there is strong two-dimensional localisation at a particular point. Note that we normalise the intensities in all figures relative to the maximum intensity for the partially-developed speckle in Fig. 2a, and note the different intensity scales used when plotting the intensity distributions. Comparing Figs. 2a and 2b, the quantitative enhancement factor of the intensity peaks in the caustic regime is 10.4, an order of magnitude. Also, we can see that the physical differences in these patterns are manifested in very different spatial spectra, with a much broader spectrum observed in Fig. 2b(iv) for the case of the caustic. The spectral width (FWHM) in this case is $\Delta k_x \sim 2.0 \, \mu m^{-1}$, consistent with a much smaller transverse peak size of $\Delta x \sim 0.5 \, \mu m$.

## IV. EXPERIMENTAL RESULTS

Our experiments were performed using identical random phase masks to those used in simulations, and intensity patterns were recorded for a range of propagation distances, allowing us to reconstruct the evolution of the field in the measurement volume shown in Fig. 1. For comparison with the simulation results shown in Fig. 2 we extracted transverse beam profiles at identical propagation distances as used in simulations (to within experimental error of ± 1 µm). These measured intensity patterns are shown in Fig. 3, with these results to be compared directly with those shown in Fig. 2 (note that for completeness we reproduce the applied phase subfigures). The experimental patterns are both normalised in the same way as in simulations, relative to the maximum intensity observed in the partially-developed speckle pattern. Note that the experimental results showing the measured pattern in the full measurement volume are displayed in animations (rotating the observation viewpoint) in Supplementary Videos S1 and S2, corresponding to the partially-developed speckle in Fig. 3(a) and the caustic regime in Fig. 3(b) respectively.

Comparing the experimental and numerical results in Fig. 2 and 3 shows excellent qualitative agreement. Note, however, that because the beam after the SLM has a random phase profile, propagation is extremely sensitive to any phase aberrations present in the experimental system, and exact quantitative agreement between the experimental and simulated patterns is not expected. Nonetheless, the experiments clearly show the same characteristic features seen in the numerical modelling – a smoothly varying intensity distribution for the partially-developed speckle and a more distinct network structure in the caustic regime. Because of the limited dynamic range in experiments (the minimum transverse peak size that could be resolved was 1.5 µm), the caustic structures in Fig 3b(i-ii) is not as apparent as that observed in simulations, but we are still able to identify the caustic network structure and distinct higher intensity peaks. Significantly, the experimentally-measured enhancement of the intensity peaks in the caustic regime is a factor of 9.6, comparable to that seen in simulations (10.4). The experimentally-measured spectral characteristics are also comparable to simulation with transverse peak sizes in the partially-developed speckle and caustic regimes of $\Delta x \sim 7.3 \, \mu m$ and $\Delta x \sim 1.5 \, \mu m$ respectively.

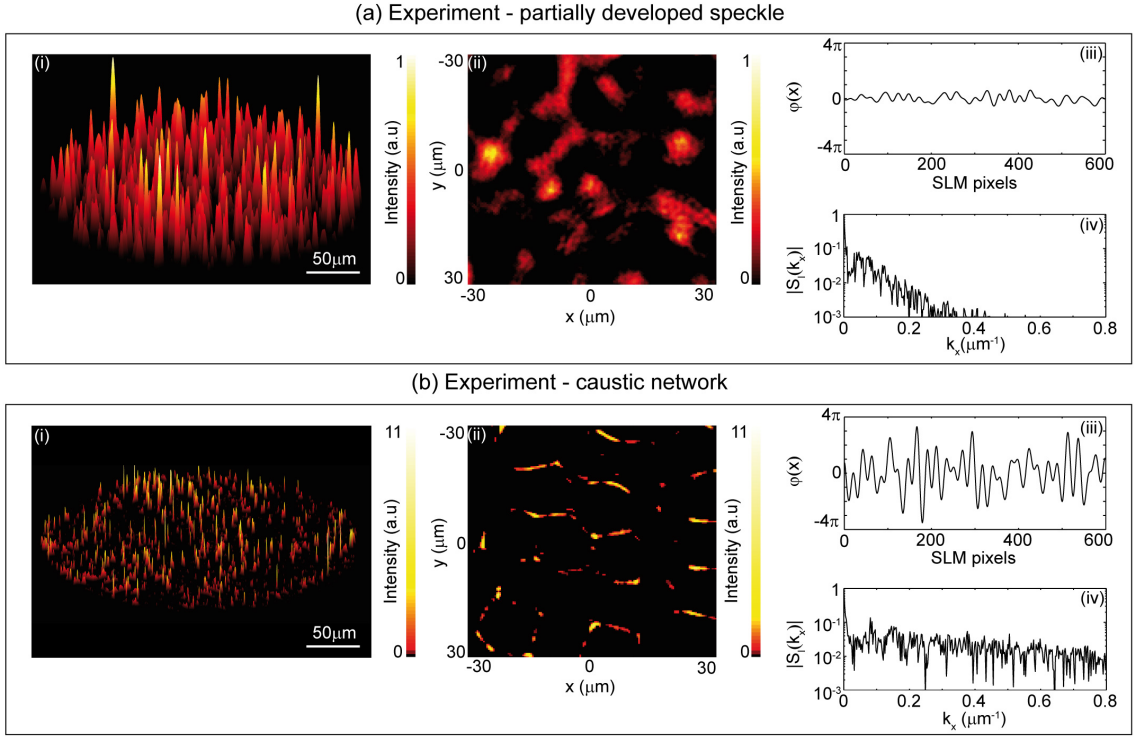

**FIGURE 3** Experimental results showing (a) partially-developed speckle and (b) a caustic network. For each case, (i) shows the computed intensity distribution; (ii) shows a zoom over a more limited region looking down on the pattern; (iii) shows a slice of the applied phase distribution to the SLM at y = 0; (iv) shows the calculated spatial spectrum corresponding to the intensity distribution in (i). Intensities in (b) are normalised relative to the maximum intensity for the partially-developed speckle in Fig. 3a. Note the different intensity scales used between (a) and (b).

## V. STATISTICAL PROPERTIES AND INTERPRETATION IN TERMS OF ROGUE WAVES

To interpret these results in the framework of optical rogue waves, it is necessary to compute the statistics of the intensity peak heights. In particular, for both the numerical and experimental results in Fig. 2 and Fig. 3, we perform two-dimensional peak detection over the spatial intensity patterns, and then compute the corresponding probability histograms, which are shown in Fig. 4. The peak detection was carried out using an 8-connected neighborhood regional maximum search, using thresholding to avoid counting the subsidiary maxima of primary peaks. Results from simulations (red asterisks) and experiments (black open circles) are compared for the cases of (a) partially-developed speckle and (b) a caustic network respectively as indicated.

Although the intensity statistics of ideal granular speckle are well-known to follow a simple exponential distribution when calculated over all field points [32], the statistics of the intensity peaks of partially-developed speckle and caustics are more complex [31,35]. For our results, what is of most interest is examining whether the intensity peak distributions show any evidence of significant long tails with events exceeding accepted criterion for rogue waves. To this end, the black and red dashed lines in each case show the calculated rogue wave intensity thresholds defined as $I_{RW} = 2I_S$, where the "significant intensity" $I_S$ is the mean of the upper third of events in the distribution [6].

On a qualitative level, the agreement between the statistics extracted from simulations and experiment for both the partially-developed speckle and the caustic regime is very good, with simulations reproducing both the general form and width of the probability distribution seen in experiment. Concerning the rogue wave criteria, simulations yield 2.3% of events in the caustic network exceeding the criterion $I_{RW}$ compared to only 0.2% for the partially-developed speckle. In experiment, we find 1.3% of events in the caustic network exceed the criterion $I_{RW}$ compared to 0.5% for the partially-developed speckle. Although the general properties of non-Gaussian statistics in speckle have been the subject of some previous studies [35-37] our results here highlight very clearly the physical connection between the appearance of caustic networks in the optical field and the formation of high intensity peaks that greatly exceed the calculated rogue wave criterion.

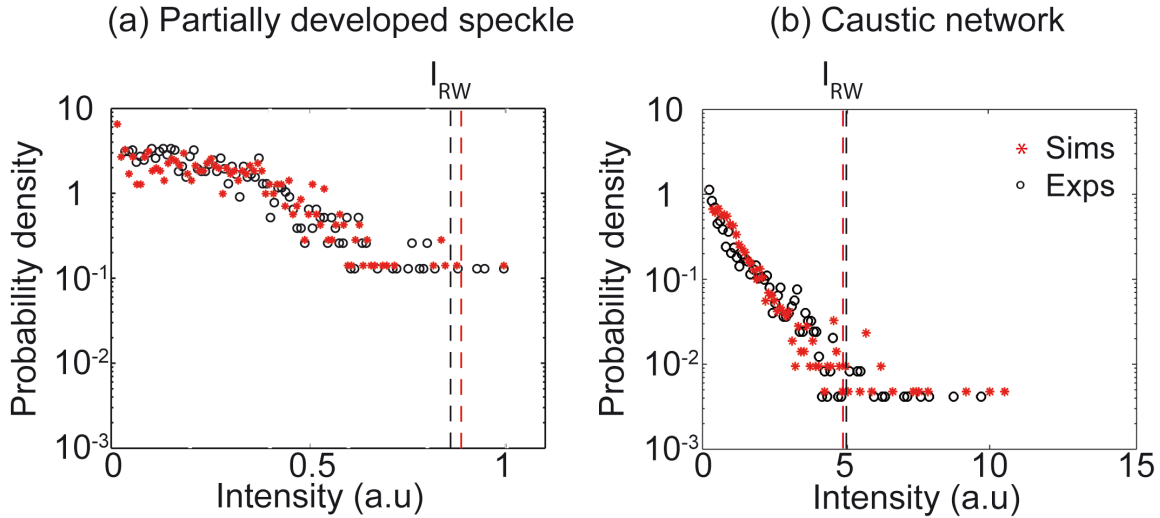

**FIGURE 4** Computed probability distributions from peak height analysis of the intensity patterns shown in Figs. 2 and 3. Results are shown both for (a) partially-developed speckle and (b) a caustic network. The red asterisks and red dashed line correspond to numerical results, the black circles and black dashed line correspond to experimental results. The label $I_{RW}$ indicates the rogue wave intensity criterion.

## VI. GENERATION OF AN "OPTICAL SEA"

The results above, as with all studies of rogue waves to date in optics, have been based on the analysis of intensity statistics. In ocean waves of course, it is not the intensity statistics that are calculated but rather the statistics of wave height, measured trough to crest relative to the zero-level of the undisplaced water surface. Significantly, although our primary experimental measurements are indeed based on an intensity pattern recorded with a CCD camera, we can use optical phase retrieval techniques to recover the associated phase and reconstruct the electric field distribution for the propagating field [38,39]. The phase retrieval technique involved recording the two dimensional intensity patterns at a number of planes in the measurement region and using the iterative Gerchberg-Saxton algorithm [38]. In our case, we recorded intensity patterns at 100 longitudinal points (spaced by 5 μm). This allows us for the first time in any optical experiment studying rogue waves to generate an amplitude distribution of an "optical sea" where we can compute wave height statistics in a way much closer to that used for ocean waves.

Our particular aim is to generate an optical field with similar statistical properties as a random sea state, which is known to have Gaussian statistics in surface elevation and Rayleigh statistics in trough to crest height [40]. Although phase retrieval on the partially-developed speckle generated with the phase in Fig. 3a(iii) yielded statistics that were very well fitted by a Gaussian distribution, to show the effect of high frequencies on extending the tails of the distribution, we modified the applied phase to yield a spatial spectrum intermediate between that of the partially-developed spectrum and the caustic network, containing high frequencies but without the strong focussing characteristics used in the generation of the caustic pattern. As we shall see this both modifies the elevation and the wave height statistics.

Fig. 5a presents our experimental results in a way similar to Fig. 3 showing intensity patterns in Fig. 5a(i) and Fig. 5a(ii), the applied phase in Fig. 5a(iii) and the spectrum in Fig. 5a(iv). The applied phase is clearly intermediate between that shown in Fig. 3a(iii) and Fig. 3b(iii), and we can indeed see how the intensity pattern shows characteristics of both the partial speckle and the caustic patterns described above. In particular, we see more strongly localised peaks than in the partially-developed speckle shown in Fig. 3a, but without the network structure of the caustic regime in Fig. 3b. Quantitatively, we can see how the peak intensity value of ~5 is also intermediate between that observed for the partially-developed speckle and the caustic network. The experimental results showing the measured pattern in the full measurement volume are displayed in an animation (rotating the observation viewpoint) in Supplementary Video S3.

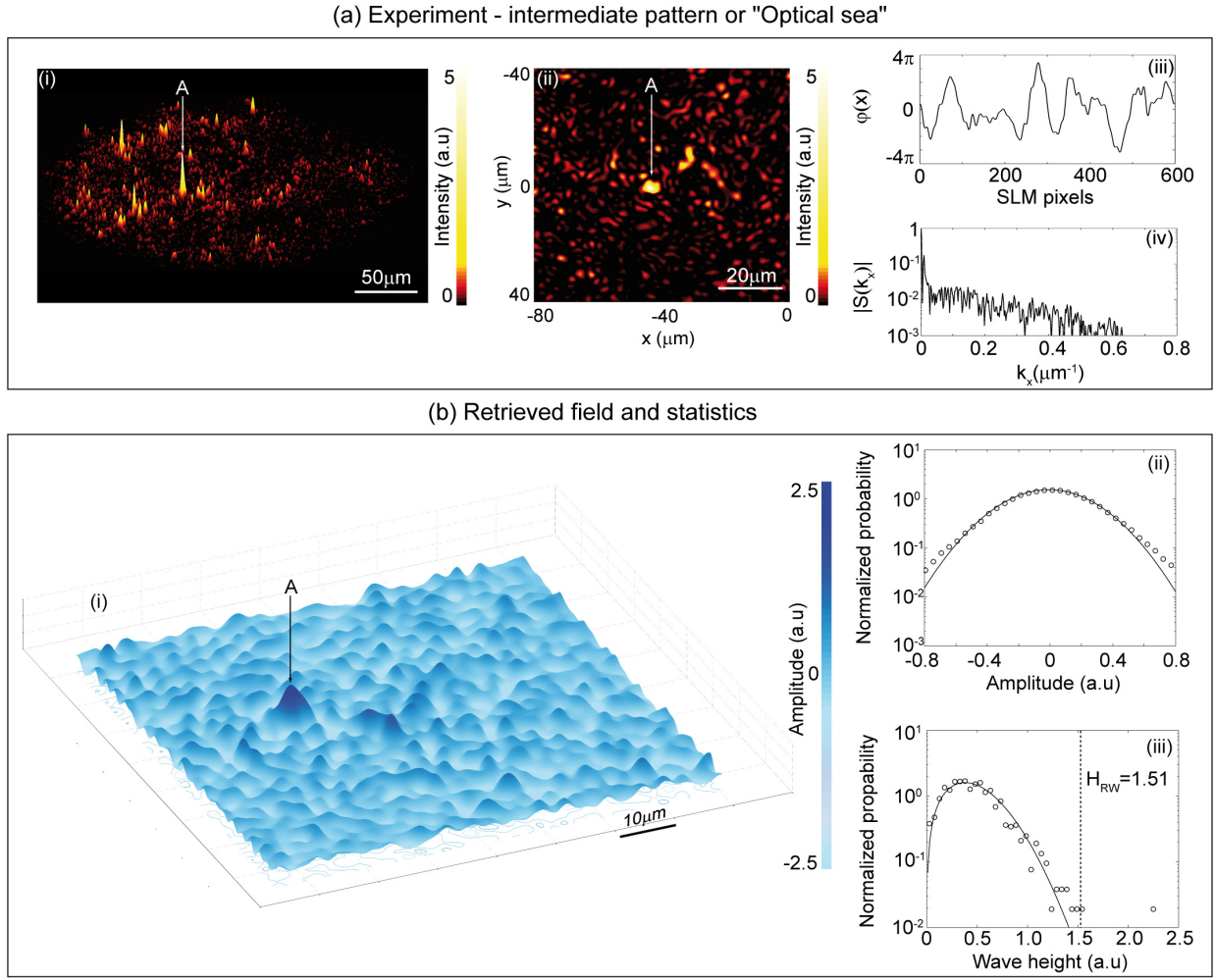

**FIGURE 5** Results obtained in a regime where phase retrieval allows experimental measurement of the spatial amplitude pattern. Fig. 5a shows the applied SLM phase, the measured intensity pattern and the corresponding spectrum at a propagation distance of 220 µm. Fig. 5b shows the retrieved amplitude pattern and computed statistics of (ii) elevation and (iii) wave height. The solid lines in these figures (ii) and (iii) plot Gaussian and Rayleigh distribution fits respectively. The bright peak A labelled in the figure is the highest peak observed in the distribution with amplitude ~ 2.23. The label $H_{RW}$ indicates the rogue wave height criterion.

Figure 5b(i) shows the retrieved field amplitude for this intensity data, which takes on both positive and negative values relative to a mean value of zero. There is clearly a striking visual resemblance to the perturbed surface of a fully-developed sea [41,42]. Based on this two-dimensional wave field, it is straightforward to determine the statistics of the field amplitude (surface elevation) and results are shown in Fig. 5b(ii). We plot the normalised probability calculated from the histogram on a log scale (circles), with the data well-fitted over the central region by a Gaussian distribution (solid line), although there are some outliers at larger elevation associated with the presence of higher spatial frequencies in the spectrum.

Based on the Gaussian statistics for elevation, we expect that the corresponding wave height distribution calculated trough to crest will be Rayleigh-distributed [1-4]. To extract wave height data from our field map, we analyse a series of wave slices parallel to the *x*-direction separated in *y* by a distance greater than the average wavelength estimated from the average zero-crossing period [43] and then measure (up-crossing) heights based on trough-to-crest distance across a zero-crossing point. Results are shown in Fig. 5b(iii) and we see the data (circles) is well-fitted by a Rayleigh distribution (solid line) aside from a significant long tail associated with the higher spatial frequencies and the presence of some rare high intensity caustic-like peaks. Note that the probability distribution fits (Gaussian, Rayleigh) shown to the data in Fig. 5b were confirmed using a Kolmogorov-Smirnov test, with the null hypothesis was accepted at the 0.05 significance level.

Indeed, the figure also plots the calculated rogue wave threshold for this data $H_{RW} = 2H_S = 1.51$ where the significant wave height $H_S$ is the mean of the upper third of events in the distribution. We see clearly how the deviation from the Rayleigh fit in the tail occurs at a wave height close to the calculated $H_S$. Note that we checked that the general features of these statistics (Gaussian elevation, Rayleigh wave height) are observed over a range of measurement distances in the experimental setup, and we also verified that they do not depend on the particular direction in which the field slices are taken in the analysis step.

## CONCLUSION

The use of a spatial light modulator to impose random phase on a coherent field has provided a flexible means of studying the rogue wave statistics of optical speckle, and to examine the particular conditions leading to extended tails in the associated probability distribution for the optical intensity. The major conclusion of our study has been to provide further evidence that a purely linear system can exhibit long tailed statistics. In particular, for the optical system under study, we identify the development of caustic structures in the random intensity pattern as the physical mechanism generating large amplitude events that satisfy commonly-applied criteria for rogue waves. Although linear wave propagation will not lead to a strongly asymmetric distribution for weakly refracting surfaces, our results in optics complement the many results from oceanography that demonstrate rogue wave formation from caustics when regions of strong focussing are present in the random medium.

We have also shown how the width of the corresponding spectrum can be correlated with the transverse size of the high intensity peaks in the pattern, again complementing studies in oceanography showing how spectral content is very important in rogue wave generation [40]. Finally, applying phase retrieval has allowed us to determine amplitude statistics for a case intermediate between a partially-developed speckle and a caustic network, with the results showing remarkable similarity to ocean wave statistics with Gaussian-distributed elevation and Rayleigh-distributed wave height. It is clear from our results that purely linear processes such as those that drive caustic focussing in wave propagation must be considered in any catalogue of mechanisms considered for rogue wave emergence.

## ACKNOWLEDGEMENTS

This work was supported by the European Research Council (ERC) Advanced Grant ERC-2011-AdG-290562 MULTIWAVE,the Agence Nationale de la Recherche (ANR OPTIROC ANR-12-BS04-0011), and the SFI (Ireland) under the programme ERC Starter Grant-Top Up, Grant 12/ERC/E2227.